\documentstyle[preprint,aps,eqsecnum]{revtex}

\begin{document}
\draft
\date{\today}
\preprint{{\Large ${\rm WU-B-94-06} \atop {\rm RUB-TPII-01/94}$}}
\title{A CRITICAL ANALYSIS OF THE PROTON FORM FACTOR WITH SUDAKOV
       SUPPRESSION AND INTRINSIC TRANSVERSE MOMENTUM
      }
\author{J. BOLZ\thanks{Supported by the Deutsche
                         Forschungsgemeinschaft}$\!$,
        R. JAKOB${}^{*}$
        and P. KROLL\footnote{E-mail address:
                               kroll@wpts0.physik.uni-wuppertal.de}$\!$
                     \thanks{Supported in part by BMFT, FRG under
                             contract 06WU737}
       }
\address{Fachbereich Physik          \\
         Universit\"at Wuppertal     \\
         D-42097 Wuppertal, Germany
         }
\author{M. BERGMANN and
        N.~G. STEFANIS${}^{*}$\footnote{E-mail address:
                               nicos@hadron.tp2.ruhr-uni-bochum.de}
       }
\address{Institut f\"ur Theoretische Physik II  \\
         Ruhr-Universit\"at Bochum              \\
         D-44780 Bochum, Germany                \\
         }
\maketitle
\newpage
\begin{abstract}
The behavior of the proton magnetic form factor is studied within the
modified hard scattering picture, which takes into account gluonic
radiative corrections in terms of transverse separations. We parallel
the analysis given previously by Li and make apparent a number of
serious objections. The appropriate cut-off needed to render the
form-factor calculation finite is both detailed and analyzed by
considering different cut-off prescriptions.
The use of the maximum interquark separation as a common infrared
cut-off in the Sudakov suppression factor is proposed, since it avoids
difficulties with the $\alpha _{s}$-singularities and yields a proton
form factor insensitive to the inclusion of the soft region which
therefore can be confidently attributed to perturbative QCD. Results are
presented for a variety of proton wave functions including also
their intrinsic transverse momentum. It turns out that the perturbative
contribution, although theoretically self-consistent for $Q^{2}$ larger
than about $6$~GeV${}^{2}$ to $10$~GeV${}^{2}$, is too small compared
to the data.
\end{abstract}
\narrowtext
\newpage 
%

\section{INTRODUCTION}
\label{sec:intro}
The proton magnetic form factor at large momentum transfer has been
extensively analyzed within perturbative Quantum Chromodynamics (pQCD)
over the last
decade~\cite{LB80,CZ84a,GS86,KS87,CGS87,JSL87,Ste89,COZ89a,Sch89,SB93a}.
The theoretical basis of these calculations is the hard scattering
formula~\cite{LB80,BF75} in which the proton form factor is
generically expressed as a convolution of a hard-scattering amplitude
$T_{H}$ and proton distribution amplitudes~(DA) $\Phi$ which represent
valence quark Fock state wave functions integrated over quark transverse
momenta (defined with respect to the momentum of their parent proton):
\begin{equation}
  G_{M}(Q^{2})
=
  \int_{0}^{1}[dx]
  \int_{0}^{1}[dx^{\prime}]
  |f_{N}(\mu _{F})|^{2}\,
  \Phi ^{\star}(x^{\prime},\mu _{F})
  T_{H}(x,x^{\prime},Q,\mu ) \Phi (x,\mu _{F}),
\label{eq:GM}
\end{equation}
where $Q^{2}$ is the invariant momentum transfer squared and
$[dx]=dx_{1}dx_{2}dx_{3}\delta (1-\sum_{}^{}x_{i})$, $x_{i}$ being the
momentum fractions carried by the valence quarks.
The renormalization scale is denoted by $\mu$ and the factorization
scale by $\mu _{F}$. The latter scale defines the interface between soft
physics---absorbed in the wave function---and hard physics, treated
explicitly within pQCD. The dimensionful constant $f_{N}$ represents
the value of the proton wave function at the origin of the configuration
space and has to be determined
nonperturbatively~\cite{CZ84a,KS87,COZ89a}. The residual (mainly
perturbative) scale dependence of $f_{N}$ and that of the proton DA is
controlled by the evolution equation~\cite{LB80}.

To lowest order the hard scattering amplitude is calculated as the sum of
all Feynman diagrams for which the three quark lines are connected
pairwise by two gluon propagators. This allows the quarks in the initial
and final proton to be viewed as moving collinearly up to transverse
momenta of order $\mu _{F}$. It is then easy to show that
$T_{H} \sim \frac{(\alpha _{s}(\mu ))^{2}}{Q^{4}}$, wherein
$\alpha _{s}$ is the running strong coupling constant in the one-loop
approximation.

The Pauli form factor $F_{2}$ and hence the electric form factor $G_{E}$
cannot be calculated within the hard scattering picture (HSP), since
they require helicity-flip transitions which are not possible for
(almost) massless quarks. These form factors are dominated by
sizeable higher twist contributions as we know from
experiment~\cite{Bos92,Kro93}. Eq.~(\ref{eq:GM}) is obtained by taking
the + component of the electromagnetic vertex and represents the
helicity-conserving part of the form factor.

The choice of the renormalization scale in the calculation of the
proton form factor is a crucial point. Most
authors~\cite{CZ84a,GS86,Ste89,COZ89a,SB93a} use a constant
$\alpha _{s}$ outside the integrals over fractional momenta,
with an argument rescaled by the characteristic virtualities for each
particular model DA. Choosing $\mu$ that way and using DAs calculated
by means of QCD sum rules---distributions whose essential characteristic
is a strong asymmetry in phase space---results for $G_{M}$ have been
obtained~\cite{COZ89a,SB93a,Ste93a} that compare fairly well with the
data~\cite{Bos92,Roc82}. On the other hand, the so-called ``asymptotic''
DA~\cite{LB80} $\Phi _{\text{as}}=120\,x_{1}x_{2}x_{3}$ ---into which
any DA should evolve with $Q^{2}\to \infty$---yields a vanishing result
for $G_{M}^{p}$.
However, for a renormalization scale independent of x, large
contributions from higher orders are expected in the endpoint region,
$x_{i}\to 0$. Indeed, for the pion form factor this has been shown
explicitly, at least for the next-to-leading order~\cite{Fie81,DR81}.
Such large higher-order contributions would render the leading-order
calculation useless. A more appropriate choice of the renormalization
scale would be, e.g.,
$\sqrt{x_{2}x^{\prime}_{2}}Q$,
since such a scale would eliminate the large logarithms arising from the
higher-order contributions. Unfortunately, this is achieved at the
expense that $\alpha _{s}$ becomes singular in the endpoint regions.
It has been conjectured~\cite{LB80} that gluonic radiative corrections
(Sudakov factors) will suppress that $\alpha _{s}$-singularity and,
therefore, in practical applications of the HSP one may handle this
difficulty by cutting off $\alpha _{s}$ at a certain value, typically
chosen in the range 0.5 to 0.7. Another, semi-phenomenological recipe to
avoid the singularity of $\alpha _{s}$ is to introduce an effective
gluon mass~\cite{Cor82} which cut-offs the interaction at low $Q^{2}$
values.

Besides the extreme sensitivity of the form factors on the utilized DA
and besides the problem with higher-order contributions and/or the
singularity of $\alpha_{s}$, there is still another---perhaps more
fundamental---difficulty with such calculations. Indeed, the
applicability of (\ref{eq:GM}) at experimentally accessible momentum
transfer, typically a few GeV, is not {\it a priori} justified. It was
argued by Isgur and Llewellyn-Smith~\cite{IL84} and also by
Radyushkin~\cite{Rad84} that the HSP receives its main contributions
from the soft endpoint regions, rendering the perturbative calculation
inconsistent. Recently, this criticism has been challenged by
Sterman and collaborators~\cite{BS89,LS92,Li93}.
Based on previous works by Collins, Soper, and Sterman~\cite{CS81}, they
have calculated Sudakov corrections to the hard-scattering process taking
into account the conventionally neglected transverse momentum,
$k_{\perp}$, of the quarks. The Sudakov corrections damp those
contributions from the endpoint regions in which transverse momenta of
the quarks are not large enough to keep the exchanged gluons hard.
Moreover, as presumed, the Sudakov corrections cancel the
$\alpha_{s}$-singularity without introducing additional {\it ad hoc}
cut-off parameters as for instance a gluon mass. Thus the modified HSP
provides a well-defined expression for the form factor which takes into
account the perturbative contributions in a self-consistent way, even
for momentum transfers as low as a few GeV.

However, an important element has not been considered in the analyses of
Refs.~\cite{LS92,Li93}. This concerns the inclusion of the intrinsic
transverse momentum of the hadronic wave function. As it was recently
shown by two of us~\cite{JK93} for the case of the pion form factor, the
inclusion of the transverse size of the pion extends considerably the
self-consistency region of the perturbative contribution down to values
of momentum transfer unreachable by the Sudakov corrections alone.
On the other hand, the incorporation of the $k_{\perp}$-dependence leads
to a substantial decrease of the magnitude of the (leading-order) pion
form factor.
Unfortunately, a clear-cut comparison with the available data is not
possible because of their low quality and the uncertainty in the
determination of the pion-nucleon coupling constant~\cite{CM90}.
Nevertheless, it seems reasonable to expect that the pion form factor
receives considerable soft contributions in the presently accessible
GeV region.

The aim of the present paper is to perform an analysis for the
proton form factor within the modified HSP. One of our objectives is to
critically examine Li's approach~\cite{Li93} and to enlarge the
theoretical framework by including the intrinsic $k_{\perp}$-dependence
of the proton wave function. At the same time we want to clarify several
technical points, which are absent in the pion case and are first
encountered in the more complicated calculation of the proton form
factor.

The purpose of our analysis is to investigate how reliably the
perturbative contribution to the proton form factor can be calculated
and to answer the question whether there is a proton wave
function---modeled on the basis of QCD sum
rules~\cite{KS87,COZ89a}---which is capable of providing, in a
theoretically self-consistent way, a good agreement with the data within
the modified HSP. It is clear that being able to identify the
leading-order perturbative contribution reliably allows us to
estimate the size of soft contributions to the proton form factor,
contributions which are not accounted for in the modified HSP formalism.
[Note that the $k_{\perp}$-dependent effects taken into account in the
modified HSP represent also soft contributions of higher-twist type.]

Sudakov suppression (which can be viewed as the perturbative
part of the transverse wave function) and intrinsic
$k_{\perp}$-dependence of the wave function may also have a lot of
interesting consequences in other exclusive reactions. Thus, for
instance, Sotiropoulos and Sterman~\cite{SS94} have applied these
elements to near-forward proton-proton elastic scattering claiming that
their interplay drives the transition of the fixed $s$ differential
cross section from the $t^{-8}$ behavior at moderate $t$ to the
$t^{-10}$ behavior at larger $t$, as predicted by dimensional counting
rules~\cite{BF73}.

The outline of the paper is as follows. In Sec.~II we discuss the proton
wave function. The modified HSP is treated in Sec.~III. The discussion
of the infrared (IR) cut-off prescription in the Sudakov factor and its
effect on the $\alpha _{s}$-singularities is given in Sec.~IV. The
numerical results are presented in Sec.~V and our conclusions are
contained in Sec.~VI.

\section{THE PROTON WAVE FUNCTION}
\label{sec:pwf}

Similarly to Sotiropoulos and Sterman~\cite{SS94}, we write the valence
quark component of the proton state with positive helicity in the form
\newpage
\begin{eqnarray}
  |P,+\!>\;
=
  \frac{1}{\sqrt{N_{c}!}}
  \int_{0}^{1}
  [dx]
  \int_{}^{}
  [d^{2}k_{\perp}]
  \Bigl\{  & \phantom{} &
\!\!\!\!\!\!
         \Psi _{123}\,{\cal M}_{+-+}^{a_{1}a_{2}a_{3}} +
         \Psi _{213}\,{\cal M}_{-++}^{a_{1}a_{2}a_{3}}
\nonumber \\
& - &
         \Bigl(\Psi _{132}\, + \,
         \Psi _{231}\Bigr){\cal M}_{++-}^{a_{1}a_{2}a_{3}}
  \Bigr\}
  \epsilon _{a_{1}a_{2}a_{3}},
\label{eq:|P,+>}
\end{eqnarray}
where we have assumed the proton to be moving rapidly in the
$3$-direction. Hence, the ratio of transverse to longitudinal momenta of
the quarks is small. The measure over the transverse momentum
integration is defined by
\begin{equation}
  [d^{2}k_{\perp}]
=
  \frac{1}{\left(16\pi ^{3}\right)^{2}} \,
  \delta ^{(2)}
               \left(
                     \sum_{i=1}^{3}\vec{k}_{\perp i}
               \right)
  d^{2}k_{\perp 1}
  d^{2}k_{\perp 2}
  d^{2}k_{\perp 3}.
\end{equation}
In the zero binding energy limit, which is characteristic for the parton
picture, one has
\begin{equation}
  x_{1} + x_{2} + x_{3} = 1
  \qquad\qquad
\hfill {\rm and} \hfill
  \qquad\qquad
  \vec{k}_{\perp 1} + \vec{k}_{\perp 2} + \vec{k}_{\perp 3} = 0.
\end{equation}

The three quark state with helicities
$
\lambda _{1}, \lambda _{2}, \lambda _{3}
$
and colors
$
a_{1}, a_{2}, a_{3}
$
is given by
\begin{equation}
  {\cal M}_{\lambda _{1}\lambda _{2}\lambda _{3}}^{a_{1}a_{2}a_{3}}
=
  \frac{1}{\sqrt{x_{1}x_{2}x_{3}}}
  |u_{a_{1}};x_{1},\vec{k}_{\perp 1},\lambda _{1}>
  |u_{a_{2}};x_{2},\vec{k}_{\perp 2},\lambda _{2}>
  |d_{a_{3}};x_{3},\vec{k}_{\perp 3},\lambda _{3}>.
\end{equation}
Since the orbital angular momentum is assumed to be zero, the proton
helicity is the sum of the quark helicities. The quark states are
normalized as follows:
\begin{equation}
  <q_{a_{i}'};x_{i}',\vec{k}_{\perp i}'\lambda _{i}'
  |q_{a^{}_{i}};x^{}_{i},\vec{k}^{}_{\perp i}\lambda ^{}_{i}> \;
=
  2x_{i}(2\pi )^{3}
  \delta _{a'_{i}a^{}_{i}}
  \delta _{\lambda _{i}'\lambda ^{} _{i}}
  \delta (x_{i}'-x^{}_{i})
  \delta (\vec{k}_{\perp i}'-\vec{k}^{}_{\perp i}).
\end{equation}
{}From the permutation symmetry between the two u quarks and from the
requirement that the three quarks have to be coupled to give an isospin
$1/2$ state it follows that Eq.~(\ref{eq:|P,+>}) can be expressed in
terms of only one independent scalar function~\cite{CZ84a}. In the
sequel, $\Psi$ denotes the momentum space wave function.

The subscripts on $\Psi$ refer to the order of momentum arguments, for
example
$
 \Psi _{123}(x,\vec{k}_{\perp})
=
 \Psi (x_{1},\vec{k}_{\perp 1};
 x_{2},\vec{k}_{\perp 2};
 x_{3},\vec{k}_{\perp 3})
$.
Note that, in general, the wave function depends on the factorization
scale $\mu _{F}$. We make the following convenient ansatz for the wave
function:
\begin{equation}
  \Psi _{123}(x,\vec{k}_{\perp})
=
  \frac{1}{8\sqrt{N_{c}!}}\,
  f_{N}(\mu _{F})
  \Phi(x,\mu _{F})\,
  \Omega (x,\vec{k}_{\perp}).
\label{eq:ansatz}
\end{equation}
The distribution amplitude $\Phi=V - A$ (in the notation
of~\cite{CZ84a}) is defined in such a way that
\begin{equation}
  \int_{0}^{1}[dx]\,
  \Phi _{123}(x,\mu _{F})
=
  1,
\label{eq:DAnorm}
\end{equation}
where an obvious abbreviated notation has been introduced.
The DA can be expressed in terms of the eigenfunctions of the evolution
equation~\cite{LB80},
$\tilde{\Phi}^{n}(x_{i})$,
which are linear combinations of Appell polynomials. Then the proton DA
can be cast into the form
\begin{equation}
  \Phi _{123}(x,\mu _{F})
=
  \Phi _{123}^{\text{as}}(x)
  \sum_{n}^{}B_{n}
  \left(\frac{\alpha _{s}(\mu _{F})}{\alpha _{s}(\mu _{0})}
  \right)^{\tilde{\gamma} _{n}/\beta_{0}}
 \tilde{\Phi}_{123}^{n}(x),
\label{eq:Phi}
\end{equation}
where the notations of~\cite{Ste89} are adopted.
$
  \Phi _{123}^{as}(x)
$
is the asymptotic DA mentioned in the Introduction. The exponents
$\tilde{\gamma}_{n}$, driving the evolution behavior of the DA, are
related to the anomalous dimensions of trilinear quark operators with
isospin $1/2$ (see~\cite{Pes79}) and resemble the $b_{n}$ in the
Brodsky-Lepage notation~\cite{LB80}. Because they are positive
fractional numbers increasing with n, higher-order terms in
(\ref{eq:Phi}) are gradually suppressed. The constants
$\tilde{\gamma}_{n}$ are given in Table~1;
$\beta_{0}=11-2n_{f}/3=9$ for three flavors.

Constraints on the DA are obtained implicitly by restricting their few
first moments within intervals determined from QCD sum
rules~\cite{CZ84a,KS87,COZ89a}, which are evaluated at some
self-consistently determined normalization point $\mu _{F}$ of order
1~GeV (see, e.g.,~\cite{Ste89}):
\begin{equation}
  \Phi ^{(n_{1}n_{2}n_{3})}(\mu _{0})
=
  \int_{0}^{1}[dx]x_{1}^{n_{1}}x_{2}^{n_{2}}x_{3}^{n_{3}}
  \Phi _{123}(x,\mu _{0})
\end{equation}
\label{eq:moments}

In most model calculations, mentioned above, the moment constraints
provided by QCD sum rules are used to determine the first five expansion
coefficients $B_{n}$, where $B_{0}=1$ due to normalization (2.7).
However, since the moments are burdened by errors, these expansion
coefficients---although mathematically uniquely determined by the
moments of corresponding order~\cite{Ste93a}---in practice their
numerical values cannot be fixed precisely giving rise to different
options for the proton DA. In our calculation of form factors we employ
amplitudes complying with the Chernyak-Ogloblin-Zhitnitsky (COZ)
sum-rule moment constraints. It was shown in \cite{BS93,BS94} that such
amplitudes constitute a finite orbit in the
$(B_{4},R\equiv |G_{M}^{n}|/G_{M}^{p})$ plane
ranging from COZ-like amplitudes~\cite{COZ89a} with $R\leq 0.5$
to the recently proposed~\cite{SB93a} heterotic one with $R\approx 0.1$.
For the convenience of the reader, the QCD sum-rules constraints and
the expansion coefficients $B_{n}$ of selected model amplitudes are
compiled in Table~1.

The $k_{\perp}$-dependence of the wave function is contained in the
function $\Omega$ which is normalized according to
\begin{equation}
  \int_{}^{}[d{}^{2}k_{\perp}]
  \Omega _{123}(x,\vec{k}_{\perp})
=
  1.
\label{eq:omeganorm}
\end{equation}
Due to (\ref{eq:DAnorm}) and (\ref{eq:omeganorm}), $f_{N}$ is the value
of the DA at the origin of the configuration space. Its evolution
behavior is given by
\begin{equation}
  f_{N}(\mu _{F})
=
  f_{N}(\mu _{0})
  \left(\frac{\alpha _{s}(\mu _{F})}{\alpha _{s}(\mu _{0})}
  \right)^{2/3\beta_{0}}
\label{eq:f_N}
\end{equation}
and its value has been determined to be
$
  f_{N}(\mu _{0})
=
  (5.0\pm 0.3)\times 10^{-3} {\rm GeV}{}^{2}
$
\cite{KS87,COZ89a}.

In Eq.~(\ref{eq:ansatz}) $\Psi $ represents the soft part of the
proton wave function, which results by removing the perturbative
part and absorbing it into the hard-scattering amplitude $T_{H}$. The
perturbative tail of the full wave function behaves as
$1/k_{\perp}^{4}$ for large $k_{\perp}$, whereas the soft part vanishes
as $1/k_{\perp}^{6}$ or faster. The nonperturbative or intrinsic
$k_{\perp}$-dependence of the soft wave function, being related to
confinement, is parametrized as a simple Gaussian according to
\begin{equation}
  \Omega _{123}(x,\vec{k}_{\perp})
=
  (16\pi ^{2})^{2}
  \frac{a^{4}}{x_{1}x_{2}x_{3}}
  exp
     \left [
            -a^{2} \sum_{i=1}^{3}k_{\perp i}^{2}/x_{i}
     \right ].
\label{eq:BLHM-Omega}
\end{equation}

This parametrization of the intrinsic $k_{\perp}$-dependence of the wave
function, which is due to Brodsky, Huang, Lepage, and
Mackenzie~\cite{BHLM83}, seems to be more favorable than the standard
form of factorizing $x$- and $k_{\perp}$-dependencies. At least for the
case of the pion wave function, this has recently been effected by
Zhitnitsky~\cite{Zhi93} on the basis of QCD sum rules. He finds that a
factorizing wave function is in conflict with some general theoretical
constraints which any reasonable wave function should comply.
Zhitnitsky's QCD sum-rule analysis of the pion wave function seems to
indicate that the $k_{\perp}$-distribution may also show a double-hump
structure, which means that small and large values of $k_{\perp}$ are
favored relative to intermediate values. It is likely that the proton
wave function may exhibit a similar behavior, though this kind of
analysis has yet to be done. For the purposes of the present work we
ignore this possibility.

In (\ref{eq:BLHM-Omega}) the parameter $a$ controls the root mean square
transverse momentum (r.m.s.), $\langle k_{\perp}^{2}\rangle ^{1/2}$, and
the r.m.s. transverse radius of the proton valence Fock state. From the
known charge radius of the proton, we expect the r.m.s. transverse
momentum to be larger than about $250$~MeV. The actual value of
$\langle k_{\perp}^{2}\rangle ^{1/2}$ may be much larger than $250$~MeV,
e.g., 600~MeV or so. Indeed, Sotiropoulos and Sterman~\cite{SS94} show
that application of the modified HSP to proton-proton elastic scattering
leads to an approximate $t^{-8}$-behavior of the differential cross
section at moderate $|t|$. The behavior $d\sigma /dt \sim t^{-10}$,
predicted by dimensional counting, appears only at very large $|t|$.
At precisely which value of $|t|$ the transition from the $t^{-8}$ to
the $t^{-10}$ behavior occurs, depends on the transverse size of the
valence Fock state of the proton. Since the ISR~\cite{ISR79} and the
FNAL~\cite{FNAL77} data are rather compatible with a $t^{-8}$-behavior
of the differential cross section, Sotiropoulos and Sterman conclude
that the transverse size of the proton is small, perhaps $\leq 0.3$ fm.
Correspondingly, the r.m.s. transverse momentum is larger than 600~MeV.
It is worth noting that such a value is supported by the findings of
the EMC group~\cite{EMC80} in a study of the transverse momentum
distribution in semi-inclusive deep inelastic $\mu p$ scattering.
A phenomenologically successful approach to the HSP, in which baryons
are viewed as bound states of a quark and an effective diquark,
also uses a value of this size for
$\langle k_{\perp}^{2}\rangle ^{1/2}$~\cite{Kro88,KPSS93,JKSS93}.
There is a second constraint on the wave function, {\it viz.} the
probability for finding three valence quarks in the proton:
\begin{equation}
  P_{3q}
=
  \frac{|f_{N}|^{2}}{3}
  (\pi a)^{4}
  \int_{0}^{1}[dx]
  \frac{2\left(\Phi _{123}(x)\right)^{2} + \Phi _{132}(x)
  \Phi _{231}(x)}{x_{1}x_{2}x_{3}}
\leq 1.
\end{equation}

In our numerical analysis to be presented in Sec.~5, we make use of two
different values of the r.m.s. transverse momentum, namely, one which is
obtained by the requirement $P_{3q}=1$ for a given wave function.
[This corresponds to the minimum value of the r.m.s. transverse
momentum.] The other option for the r.m.s. transverse momentum we
consider is the rather large value of 600~MeV. In the latter case, the
probability for the valence quark Fock state depends on the wave
function.

\section{THE MODIFIED HARD SCATTERING PICTURE}
\label{sec:mhsp}

Following Li~\cite{Li93}, we write the proton form factor in the form
\begin{equation}
  G_{\text{M}}(Q^{2})
=
  \frac{16}{3}
  \int_{0}^{1}[dx][dx']
  \int_{}^{}[d{}^{2}k_{\perp}][d{}^{2}k_{\perp}^{\prime}]
  \sum_{j=1}^{2}
  T_{H_{j}}(x,x',\vec{k}_{\perp},\vec{k}'_{\perp},Q,\mu )
  Y_{j}(x,x',\vec{k}_{\perp},\vec{k}'_{\perp},\mu _{F}).
\label{eq:G_M}
\end{equation}
Note, however, that our notation is slightly different compared to
that of Li.
Making use of the symmetry properties of the proton wave function under
permutation, the contributions from the 42 diagrams involved in the
calculation of the proton form factor in lowest order can be arranged
into two reduced hard scattering amplitudes of the form
\begin{equation}
  T_{H_{1}}
=
  \frac{2}{3}\,C_{\text{F}}\,
  \frac{(4\pi \alpha _{s}(\mu ))^{2}}
       {
        \left[
                (1-x_{1})(1-x_{1}')Q^{2}
              + (\vec{k}_{\perp 1} - \vec{k}_{\perp 1}')^{2}
        \right]
        \left[
                x_{2}x_{2}'Q^{2}
              + (\vec{k}_{\perp 2} - \vec{k}_{\perp 2}')^{2}
        \right]
       },
\label{eq:T_H_1}
\end{equation}
\begin{equation}
  T_{H_{2}}
=
  \frac{2}{3}\,C_{\text{F}}\,
  \frac{(4\pi \alpha _{s}(\mu ))^{2}}
       {
        \left[
                x_{1}x_{1}'Q^{2}
              + (\vec{k}_{\perp 1} - \vec{k}_{\perp 1}')^{2}
        \right]
        \left[
                x_{2}x_{2}'Q^{2}
              + (\vec{k}_{\perp 2} - \vec{k}_{\perp 2}')
        \right]
       },
\label{eq:T_H_2}
\end{equation}
where $C_{\text{F}}=4/3$ is the Casimir operator of the fundamental
representation of $SU(3)_{c}$. In the hard scattering amplitudes only
the $k_{\perp}$-dependence of the gluon propagators is included, whereas
that of the quark propagators has been neglected. It is expected that
this technical simplification introduces only a minor error of about
$10\%$ in the final result. For the case of the pion form factor this
has been explicitly demonstrated by Li~\cite{Li93}.

The functions $Y_{j}$ in (\ref{eq:G_M}) are short-hand notations for
linear combinations of products of the initial and final state wave
functions
$
 \Psi _{ijk}\Psi ^{\star\prime} _{i^{\prime}j^{\prime}k^{\prime}}
$,
weighted by $x_{i}$-dependent factors arising from the fermion
propagators, namely:
\begin{eqnarray}
  Y_{1}
=
  \frac{1}{(1-x_{1})(1-x^{\prime}_{1})}
\Bigl\{  & \phantom{} &
\!\!\!\!\!\!
           4\Psi ^{\star\prime}_{123}\Psi _{123}
         + 4\Psi ^{\star\prime}_{132}\Psi _{132}
         +  \Psi ^{\star\prime}_{231}\Psi _{231}
         +  \Psi ^{\star\prime}_{321}\Psi _{321}
\nonumber \\
&  +  &    2\Psi ^{\star\prime}_{231}\Psi _{132}
         + 2\Psi ^{\star\prime}_{132}\Psi _{231}
         + 2\Psi ^{\star\prime}_{321}\Psi _{123}
         + 2\Psi ^{\star\prime}_{123}\Psi _{321}
  \Bigr\}
\label{eq:Y_1}
\end{eqnarray}
\begin{eqnarray}
  Y_{2}
= & \phantom{} &
\!\!\!\!\!\!
  \frac{1}{2(1-x_{2})(1-x^{\prime}_{1})}
  \left\{
           3\Psi ^{\star\prime}_{132}\Psi _{132}
         -  \Psi ^{\star\prime}_{231}\Psi _{231}
         -  \Psi ^{\star\prime}_{231}\Psi _{132}
         -  \Psi ^{\star\prime}_{132}\Psi _{231}
  \right\}
\nonumber \\
& - & \frac{1}{(1-x_{3})(1-x^{\prime}_{1})}
  \left\{
           4\Psi ^{\star\prime}_{321}\Psi _{321}
         +  \Psi ^{\star\prime}_{123}\Psi _{123}
         + 2\Psi ^{\star\prime}_{321}\Psi _{123}
         + 2\Psi ^{\star\prime}_{123}\Psi _{321}
  \right\}.
\label{eq:Y_2}
\end{eqnarray}

Ignoring the transverse momenta in the hard scattering amplitudes
(\ref{eq:T_H_1}) and (\ref{eq:T_H_2}), and inserting (\ref{eq:ansatz})
and (\ref{eq:omeganorm}), one arrives at the standard HSP result for the
magnetic form factor. Although this expression is correct in the
asymptotic momentum domain, the transverse degrees of freedom are an
essential ingredient of the formalism and neglecting them leads to
inconsistencies in the endpoint regions, where one of the fractional
momenta $x_{i}$ or $x^{\prime}_{i}$ tends to zero. After all, it is
precisely this approximation that is responsible for the
inconsistencies mentioned in the Introduction. The power of combining
the transverse momentum dependence of the hard scattering amplitude and
radiative corrections in the form of Sudakov form factors was realized
by Sterman and collaborators~\cite{BS89,LS92,Li93}. Ultimately, it leads
to a suppression of contributions from the dangerous soft regions, where
both the longitudinal and transverse momenta of the quarks are small.

In order to include the Sudakov corrections, it is advantageous to
reexpress Eq.(~\ref{eq:G_M}) in terms of the variables $\vec{b}_{i}$,
which are canonically conjugate to $\vec{k}_{\perp i}$ and span the
transverse configuration space. Then
\begin{equation}
  G_{\text{M}}(Q^{2})
=
  \frac{16}{3}
  \int_{0}^{1}[dx][dx']
  \int_{}^{}\frac{d{}^{2}b_{1}}{(4\pi )^{2}}
            \frac{d{}^{2}b_{2}}{(4\pi )^{2}}
  \sum_{j}^{}\, \hat{T}_{j}(x,x',\vec{b},Q,\mu )
  \hat{Y}_{j}(x,x',\vec{b},\mu _{F}) \,
  {\rm e}^{-S_{j}},
\label{eq:G_M(b)}
\end{equation}
where the Fourier transform of a function
$f(\vec{k}_{\perp})=f(\vec{k}_{\perp 1},\vec{k}_{\perp 2})$
is defined by
\begin{equation}
  \hat{f}(\vec{b})
=
  \frac{1}{(2\pi )^{4}}
  \int_{}^{}d{}^{2}k_{\perp 1}d{}^{2}k_{\perp 2}
  {\rm exp}\{-i\vec{b}_{1}\!\!\cdot \vec{k}_{\perp 1}
 - i\vec{b}_{2}\!\cdot \vec{k}_{\perp 2}\}
  f(\vec{k}).
\label{eq:Fourier}
\end{equation}

Since the hard scattering amplitudes depend only on the differences of
initial and final state transverse momenta, there are only two
independent Fourier-conjugate vectors
$\vec{b}_{1}\;(=\vec{b}_{1}')$
and
$\vec{b}_{2}\;(=\vec{b}_{2}')$. They are, respectively,
the transverse separation vectors between quarks 1 and 3 and between
quarks 2 and 3. Accordingly, the transverse separation of quark 1 and
quark 2 is given by
\begin{equation}
  \vec{b}_{3}
=
  \vec{b}_{2} - \vec{b}_{1}.
\label{eq:b123}
\end{equation}
[Note that Sotiropoulos and Sterman~\cite{SS94} define the transverse
separations in a cyclic way which results in the interchange
$\vec{b}_{1} \longleftrightarrow -\vec{b}_{2}$, as compared to our
definition.]

The fact that there are only two independent transverse separation
vectors is a consequence of the approximation made in the treatment
of the hard scattering amplitudes (\ref{eq:T_H_1}) and (\ref{eq:T_H_2})
which disregards the $k_{\perp}$-dependence of the quark propagators.
This approximation is justified by the enormous technical simplification
it entails, given that the thereby introduced errors are very small.
Then by virtue of rotational invariance of the system with respect to
the longitudinal axis, the form factor (\ref{eq:G_M(b)}) can be expressed
in terms of a seven-dimensional integral instead of an eleven-dimensional
one. Physically, the relations
$
 \vec{b}_{1}=\vec{b}_{1}'
$,
$
 \vec{b}_{2}=\vec{b}_{2}'
$
mean that the physical probe (i.e., the photon) mediates only such
transitions from the initial to the final proton state, which have the
same transverse configurations of the quarks.

The Fourier-transformed hard scattering amplitudes appearing in
Eq.~(\ref{eq:G_M(b)}) read
\begin{equation}
  \hat{T}_{1}
=
  \frac{8}{3}\,C_{\text{F}}\,\alpha _{s}(t_{11}) \alpha _{s}(t_{12})
  K_{0}
       \left(
             \sqrt{(1-x_{1})(1-x_{1}')}Qb_{1}
       \right)
  K_{0}\left(
             \sqrt{x_{2}x_{2}'}Qb_{2}
       \right),
\label{eq:FourierT_1}
\end{equation}
\begin{equation}
  \hat{T}_{2}
=
  \frac{8}{3}\,C_{\text{F}}\,\alpha _{s}(t_{21}) \alpha _{s}(t_{22})
  K_{0}
       \left(
             \sqrt{x_{1}x_{1}'}Qb_{1}
       \right)
  K_{0}
       \left(
             \sqrt{x_{2}x_{2}'}Qb_{2}
       \right),
\label{eq:FourierT_2}
\end{equation}
where $K_{0}$ is the modified Bessel function of order 0 and $b_{l}$
denotes the length of the corresponding vector.
We have now chosen the renormalization scale in such a way that each
hard gluon carries its own individual momentum scale $t_{ji}$ as
the argument of the corresponding $\alpha _{s}$.
The $t_{ji}$ are defined as the maximum scale of either the longitudinal
momentum or the inverse transverse separation, associated with each of
the gluons:
\begin{eqnarray}
& t_{11} &
=
  {\rm max} \left[
                  \sqrt{(1-x_{1})(1-x_{1}^{\prime})Q}, 1/b_{1}
            \right],
\nonumber \\
& t_{21} &
=
  {\rm max} \left[
                  \sqrt{x_{1}x_{1}^{\prime}Q}, 1/b_{1}
            \right],
\nonumber \\
& t_{12} &
=
  t_{22}
=
  {\rm max} \left[
                  \sqrt{x_{2}x_{2}^{\prime}Q}, 1/b_{2}
            \right],
\label{eq:t_ij}
\end{eqnarray}
One may think of other choices. However, they are not expected to lead
to very different predictions for the form factor~\cite{Li93}.

The quantities $\hat{Y}_{j}$ contain the same combinations
of initial and final state wave functions as those in (\ref{eq:Y_1}) and
(\ref{eq:Y_2}), the only difference being that now the products
$
 \Psi ^{\star\prime}_{i^{\prime}j^{\prime}k^{\prime}}
 \Psi _{ijk}
$
are replaced by corresponding products of Fourier-transformed wave
functions:
$
 \hat{\Psi}^{\star\prime}_{i^{\prime}j^{\prime}k^{\prime}}
           (x',\vec{b},\mu _{F})
 \hat{\Psi}_{ijk}(x,\vec{b},\mu _{F})
$.
Using (\ref{eq:ansatz}) and (\ref{eq:BLHM-Omega}), the Fourier transform
of the wave function reads
\begin{equation}
  \hat{\Psi}_{123}(x,\vec{b},\mu _{F})
=
  \frac{1}{8\sqrt{N_{\text{c}}!}}
  f_{\text{N}}(\mu _{F})
  \Phi _{123}(x,\mu _{F})
  \hat{\Omega}_{123}(x,\vec{b}),
\label{eq:FourierPsi}
\end{equation}
where the Fourier-transform of the $k_{\perp}$-dependent part is given by
\begin{equation}
  \hat{\Omega}_{123}(x,\vec{b})
=
  (4\pi )^{2}
  {\rm exp}
           \left\{
                  - \frac{1}{4a^{2}}
                  \Bigl[
                          x_{1}x_{3}b_{1}^{2} + x_{2}x_{3}b_{2}^{2}
                        + x_{1}x_{2}b_{3}^{2}
                  \Bigr]
           \right\}.
\label{eq:FourierOmega}
\end{equation}

The exponentials ${\rm e}^{-S_{j}}$ in (\ref{eq:G_M(b)}) are the
Sudakov factors, which incorporate the effects of gluonic radiative
corrections. Because of this, (\ref{eq:G_M(b)}) is not simply the
Fourier transform of (\ref{eq:G_M}) but an expression comprising
an additional physical input. Thus (\ref{eq:G_M(b)}) may be termed the
``modified hard-scattering formula''.
On the ground of previous works by Collins and Soper~\cite{CS81}, Botts
and Sterman~\cite{BS89} have calculated a Sudakov factor using
resummation techniques and having recourse to the renormalization group.
They find Sudakov exponents of the form
\begin{eqnarray}
  S_{j}
= & \phantom{} &
\!\!\!\!\!\!\!
  \sum_{l=1}^{3}
  \left[
          s(x_{l},\tilde{b}_{l},Q)
        + \int_{1/\tilde{b}_{l}}^{t_{j1}}
                    \frac{d\bar{\mu}}{\bar{\mu}}
                    \gamma _{q}(g(\bar{\mu} ^{2}))
  \right]
\nonumber \\
& + &
  \sum_{l=1}^{3}
  \left[
          s(x_{l}^{\prime},\tilde{b}_{l},Q)
        + \int_{1/\tilde{b}_{l}}^{t_{j2}}
                    \frac{d\bar{\mu}}{\bar{\mu}}
                    \gamma _{q}(g(\bar{\mu} ^{2}))
  \right],
\label{eq:S}
\end{eqnarray}
wherein the Sudakov functions $s(\xi _{l},\tilde{b}_{l},Q)$ are given by
\begin{eqnarray}
  s(\xi _{l},\tilde{b}_{l},Q)
= & \phantom{} &
\!\!\!\!\!\!\!\!
   \frac{A^{(1)}}{2\beta _{1}}\;
   \hat{q}_{l}
   {\rm ln}\!\left(
                 \frac{\hat{q}_{l}}{\hat{b}_{l}}
           \right)
 +
   \frac{A^{(2)}}{4\beta _{1}^{2}}\,
           \left(
                 \frac{\hat{q}_{l}}{\hat{b}_{l}} - 1
           \right)
 - \frac{A^{(1)}}{2\beta _{1}}\;
            (\hat{q}_{l}-\hat{b}_{l})
\nonumber \\
& - &
   \frac{A^{(1)}\beta _{2}}{16\beta _{1}^{3}}\,
   \hat{q}_{l}
           \left[
                   \frac{{\rm ln}(2\hat{b}_{l}) + 1}{\hat{b}_{l}}
                 -
                   \frac{{\rm ln}(2\hat{q}_{l}) + 1}{\hat{q}_{l}}
           \right]
\nonumber \\
& - &
  \left[
          \frac{A^{(2)}}{4\beta _{1}^{2}}
        - \frac{A^{(1)}}{4\beta _{1}}
                   {\rm ln}\Bigl({\rm e}^{2\gamma -1}/2\Bigr)
  \right]
  {\rm ln}\left(
                \frac{\hat{q}_{l}}{\hat{b}_{l}}
          \right)
\nonumber \\
& - &
  \frac{A^{(1)}\beta _{2}}{32\beta _{1}^{3}}
           \left[
                   {\rm ln}^{2}(2\hat{q}_{l})
                 - {\rm ln}^{2}(2\hat{b}_{l})
           \right].
\label{eq:s}
\end{eqnarray}
Here
$\xi _{l}=x_{l}$ or $x_{l}^{\prime}$ ($l=1,2,3$)
and the variables $\hat{q}$ and $\hat{b}$ are defined as follows:
\begin{equation}
  \hat{q}_{l}
=
  {\rm ln}[\xi _{l}Q/(\sqrt{2}\Lambda _{\text{QCD}})]
\end{equation}
\begin{equation}
  \hat{b}_{l}
=
  {\rm ln}[1/\tilde{b}_{l}\Lambda _{\text{QCD}}].
\end{equation}
The coefficients $A^{(i)}$ and $\beta _{i}$ are
\[
  A^{(1)} = \frac{4}{3}, \;\;\;
  A^{(2)} =   \frac{67}{9} - \frac{1}{3}\pi ^{2} - \frac{10}{27}n_{f}
            + \frac{8}{3}\beta _{1}{\rm ln}\!
              \left(
                    \frac{1}{2}{\rm e}^{\gamma}
              \right),
\]
\begin{equation}
  \beta _{1}=\frac{33-2n_{f}}{12},\;\;\;\;\;
  \beta _{2}=\frac{153-19n_{f}}{24},
\end{equation}
where $n_{f}$ is the number of quark flavors and $\gamma =0.5772\ldots$
is the Euler-Mascheroni constant. In the sequel $n_{f}=3$ is used.
$\gamma _{q}=-\frac{\alpha _{s}}{\pi}\;+\;{\rm O}(\alpha _{s}^{2})$ is
the anomalous quark dimension in the axial gauge~\cite{Sop80}.

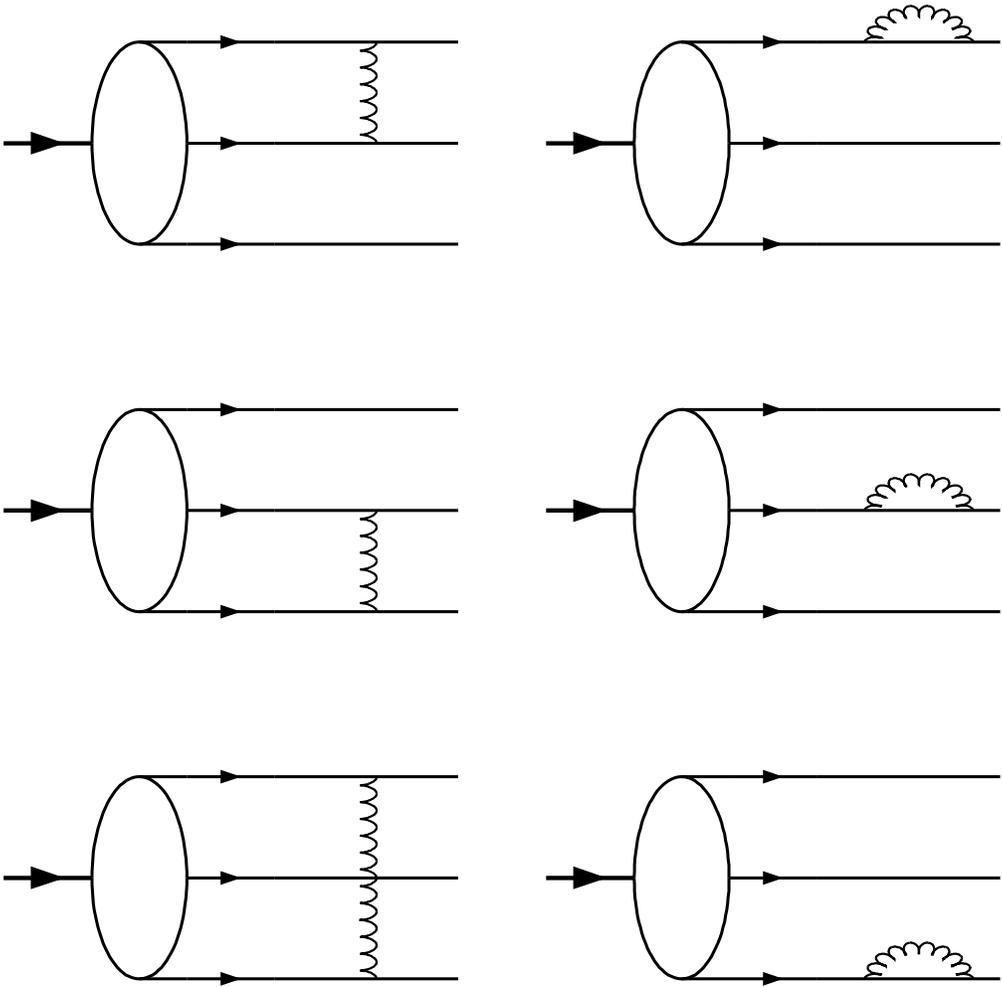
\begin{figure}
\caption{Illustration of gluonic radiative corrections to the proton
         magnetic form factor in the axial gauge.}
\label{fig:gluonlines}
\end{figure}

The Sudakov function, $s(\xi _{l},\tilde{b}_{l},Q)$, in~(\ref{eq:s})
takes into account leading and next-to-leading gluonic radiative
corrections of the form shown in Fig.~\ref{fig:gluonlines}.
The quantities $\tilde{b}_{l}$ ($l=1,2,3$) are infrared cut-off
parameters, naturally related to, but not uniquely determined by the
mutual separations of the three quarks~\cite{CS81}.
A physical perspective on the choice of the IR cut-off is provided
by the following analogy to ordinary QED. One expects that because of
the color neutrality of a hadron, its quark distribution cannot be
resolved by gluons with a wave length much larger than a characteristic
quark separation scale; meaning that long wave length gluons probe the
color singlet proton and hence radiation is damped.
Radiative corrections with wave lengths between the IR cut-off and an
upper limit (related to the physical momentum Q) yield to suppression;
it is understood that still softer gluonic corrections are already
taken care of in the hadron wave function, whereas harder gluons are
considered as part of $T_{H}$.

Different choices of the IR cut-off have been used in the literature:
Thus, Li~\cite{Li93} chooses $\tilde{b}_{l}=b_{l}$ (this choice
hereafter is termed the ``L'' prescription), whereas
Hyer~\cite{Hye93} in his analysis of the proton-antiproton
annihilation into two photons and of the time-like proton form factor as
well as Sotiropoulos and Sterman~\cite{SS94} take
$\tilde{b}_{1}=b_{2}$, $\tilde{b}_{2}=b_{1}$, $\tilde{b}_{3}=b_{3}$
(this choice is denoted the ``H-SS'' prescription).
Still another possibility, and the one proposed in the present work for
reasons that will be explained below is to use as IR cut-off the maximum
of the three interquark separations, i.e., to set
\begin{equation}
  \tilde{b}\equiv {\rm max}\{b_{1},b_{2},b_{3}\}
=
  \tilde{b}_{1}=\tilde{b}_{2}=\tilde{b}_{3}.
\label{eq:MAX}
\end{equation}

This choice, designated by ``MAX'', is analogous to that in the meson
case, wherein the quark-antiquark distance naturally provides a secure
IR cut-off. The specific features of each particular cut-off choice will
be discussed in detail in Sec.~\ref{sec:singularities}.

The integrals in~( \ref{eq:S}) arise from the application of the
renormalization group equation (RGE). The evolution from one scale
value to another is governed by the anomalous dimensions of the involved
operators. The integrals combine the effects of the application
of the RGE on the wave functions and on the hard scattering amplitude.
The range of validity of (\ref{eq:s}) for the Sudakov
functions is limited to not too small $\tilde{b_{l}}$ values. Whenever
$1/\tilde{b}_{l}$ is large relative to the hard (gluon) scale
$\xi _{l}Q$, the gluonic corrections are to be considered as
higher-order corrections to $T_{H}$ and hence are not contained in the
Sudakov factor but are absorbed in $T_{H}$. For that reason,
Li~\cite{Li93} sets any Sudakov function $s(\xi _{l},\tilde{b}_{l},Q)$
equal to zero whenever $\xi _{l}\leq \sqrt{2}/(Q\tilde{b}_{l})$.
Moreover, Li holds the Sudakov factor ${\rm e}^{-S_{j}}$ equal to unity
whenever it exceeds this value, which is the case in the small
$\tilde{b}_{l}$-region.
Actually, the full expression~(\ref{eq:S}) shows in this region a small
enhancement resulting from the interplay of the next-to-leading
logarithmic contributions to the Sudakov exponents and the integrals
over the anomalous dimensions. We follow the same lines of argument in
our analysis.

The IR cut-offs $1/\tilde{b}_{l}$ in the Sudakov exponents mark the
interface between the nonperturbatively soft momenta, which are
implicitly accounted for in the proton wave function, and the
contributions from soft gluons, incorporated in a perturbative way in
the Sudakov factors. Obviously, the IR cut-off serves at the same time
as the gliding factorization scale $\mu _{F}$ to be used in the
evolution of the wave function. For that reason, Li~\cite{Li93} as well
as Sotiropoulos and Sterman~\cite{SS94} take
$\mu _{F} = {\rm min}\{1/\tilde{b}_{l}\}$.
The ``MAX'' prescription (\ref{eq:MAX}), adopted in the present work,
naturally complies with the choice of the evolution scale proposed
in~\cite{Li93,SS94}.

\section{DISCUSSION OF THE $\lowercase{\alpha _{s}}$-SINGULARITIES}
\label{sec:singularities}
It is well known that the inclusion of an $x$-dependent renormalization
scale in the argument of $\alpha _{s}$ within the standard HSP of
Brodsky-Lepage~\cite{LB80} presents the difficulty that the value of
$\alpha _{s}$ becomes singular in the endpoint regions. To render the
form factors (Eq.~(\ref{eq:GM})) finite, additional external
parameters, like an effective gluon mass~\cite{Cor82} or a cut-off
prescription have to be introduced. Technically, such parameters play
the r\^{o}le of IR regulators serving to regularize one of the
gluon propagators, which may become soft along the boundaries of phase
space (see, e.g., ~\cite{Ste89}). One of the crucial advantages of the
modified HSP, proposed by Sterman and collaborators~\cite{BS89,LS92,Li93},
is that there is no need for external regulators because the Sudakov
factor may suppress the singularities of the ``bare'' (one-loop) $\alpha
_{s}$ inherently.
Indeed in the pion case, it was shown~\cite{LS92} that the transverse
quark-antiquark separation is tantamount to an IR regulator which
suffices to cancel all singularities from the soft region.

Concerning the proton form factor, the situation is much more complicated
because more scales are involved and hence the choice of the appropriate
IR cut-off parameters $\tilde{b}_{l}$ is not obvious, as discussed in
Sec.~III.
As we shall effect in the following, the cancellation of the
$\alpha _{s}$-singularities by the Sudakov factor depends
sensitively on that particular choice.

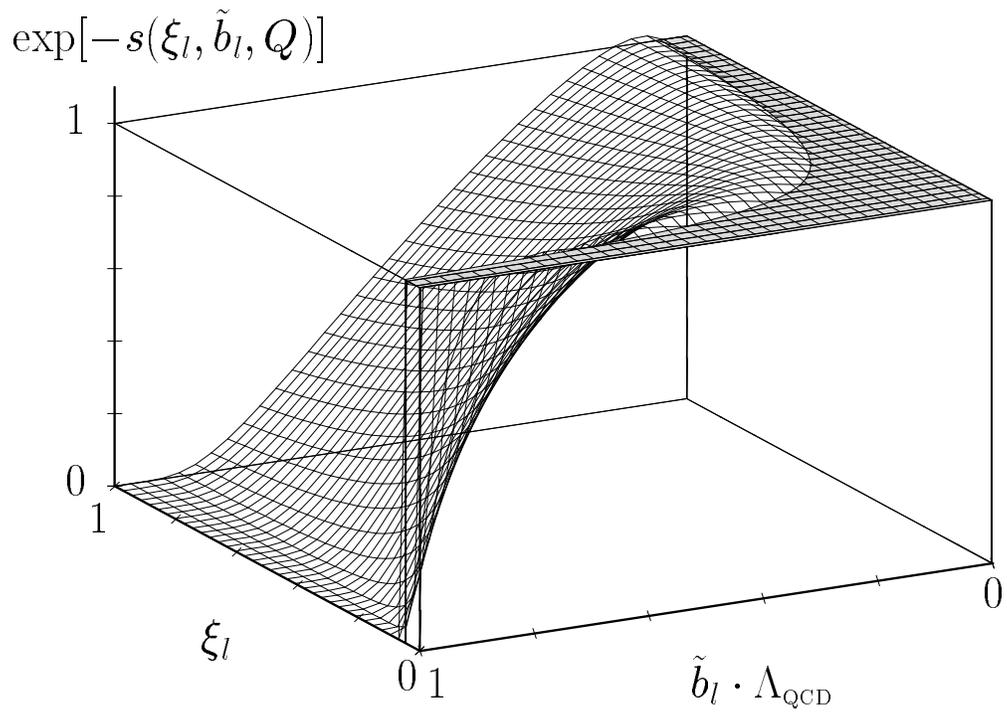
\begin{figure}
\caption[fig:sud]{The exponential of the Sudakov function
         $s(\xi _{l},\tilde{b}_{l},Q)$ vs. $\xi _{l}$ and
         $\tilde{b}_{l}\Lambda _{\text{QCD}}$ for
         $Q=30 \Lambda _{\text{QCD}}$. In the hatched area the Sudakov
         function is set equal to zero according to Li's requirement
         \cite{Li93}.}
\label{fig:sudakov}
\end{figure}

In Fig.~\ref{fig:sudakov} we display the exponential of the Sudakov
function
$\exp[-s(\xi _{l},\tilde{b}_{l},Q)]$ for $Q=30\: \Lambda _{\text{QCD}}$
by imposing Li's requirement~\cite{Li93}:
$s(\xi _{l},\tilde{b}_{l},Q)=0$ whenever
$\xi _{l}\leq \sqrt{2}/Q\tilde{b}_{l}$.
Ultimately, the cancellation of the $\alpha _{s}$-singularities relies
on the fact that whenever one of the $\alpha _{s}$ tends to infinity
(owing to the limit $t_{ji}\to \Lambda _{\text{QCD}}$), the Sudakov
factor ${\rm e}^{-S_{j}}$ rapidly decreases to zero. As it can be
observed from Fig.~\ref{fig:sudakov} this is not the case in the region
determined by
$\xi _{l}\le \sqrt{2}\Lambda _{\text{QCD}}/Q$ and simultaneously
$\tilde{b}_{l}\Lambda _{\text{QCD}}\to 1$, where
$\exp[-s(\xi _{l},\tilde{b}_{l},Q)]$ is fixed to unity. In the pion case
this does not matter, since the other
$\exp[-s(1-\xi ,\tilde{b},Q)]\to 0$ faster than any power of
$\ln[1/(\tilde{b}\Lambda _{\text{QCD}})]$ and,
consequently, the Sudakov factor drops to zero. In contrast,
the treatment of the proton form factor is more subtle. In that case,
${\rm e}^{-S_{j}}$ does not necessarily vanish fast enough to guarantee
the cancellation of the $\alpha _{s}$-singularities.
This can be illustrated by the following configuration: if, say,
$x_{1}< \sqrt{2}\Lambda _{\text{QCD}}/Q$ and
$\tilde{b}_{1}\Lambda _{\text{QCD}}\to 1$
then $x_{2}+x_{3}\approx 1$ and $x_{2}$ can have any value
between 0 and $1-x_{1}$. Since $\tilde{b}_{2}\Lambda _{\text{QCD}}$
is unrestricted within the limits $0$ and $1$, the corresponding
exponentials of the Sudakov functions
$\exp[-s(x_{2},\tilde{b}_{2},Q)]$ and
$\exp[-s(x_{3},\tilde{b}_{3},Q)]$ do not
automatically fall off to zero in order to yield sufficient suppression
of the $\alpha _{s}$-singularities, unless all three $\tilde{b}_{l}$
are coerced to be equal.
If the three $\tilde{b}_{l}$ are allowed to be different, then the
Sudakov factor provides suppression only through the contributions of
the anomalous dimensions. According to the ``L'' and ``H-SS''
prescriptions, which, in general, allow for different $\tilde{b}_{l}$ in
the Sudakov functions, the integrand in (\ref{eq:G_M(b)}) has
singularities behaving as
\begin{equation}
  \sim \ln \left(
                 \frac{1}{\tilde{b}_{l}\Lambda _{\text{QCD}}}
           \right)^{\kappa}
\label{eq:sing}
\end{equation}
for $\tilde{b}_{l}\Lambda _{\text{QCD}} \simeq 1$ and $x_{l}$ hold
fixed. The maximum degree of divergence is given by
\begin{equation}
  \kappa
=
  \frac{1}{\beta _{0}}
                      \left(
                            \frac{4}{3} + 2\tilde{\gamma}_{\text{max}}
                             - 2
                      \right)
                              + 1,
\label{eq:kappa}
\end{equation}
where the first term $4/3$ comes from the evolution of $f_{N}$,
(\ref{eq:f_N}) and the constant $\tilde{\gamma}_{\text{max}}$
is related to the anomalous dimension driving the evolution behavior of
the proton DA, see (\ref{eq:Phi}) and Table~1:
$\tilde{\gamma}_{\text{max}}$ is the maximum value of the
$\{\tilde{\gamma}_{n}\}$ within a given polynomial order of the
expansion of the DA.
We reiterate that the $\tilde{\gamma}_{n}$ are positive fractional
numbers increasing with $n$. Thus the singular behavior of the integrand
becomes worse as the expansion in terms of Appell polynomials extends to
higher and higher orders. The term $-2$ in (\ref{eq:kappa}) stems from
the integrations over the anomalous dimensions in the Sudakov factor
${\rm e}^{-S_{j}}$ (see (\ref{eq:S})). Finally, the term $1$
originates from that $\alpha _{s}(t_{jk})$ which becomes singular
in (\ref{eq:G_M(b)}), c.f., (\ref{eq:t_ij}). Which one of the
$\alpha _{s}$ couplings becomes actually singular, depends on the
prescription imposed on the IR cut-off parameters $\tilde{b}_{l}$.
The integral (\ref{eq:G_M(b)}) does not exist if
$\tilde{\gamma}_{\text{max}}\geq \frac{1}{3}$.
As Table~1 reveals, this happens already for proton DAs  which include
Appell polynomials of order 1, i.e., for all DAs except for the
asymptotic one: $\Phi _{\text{as}}=120\,x_{1}x_{2}x_{3}$. Thus
application of the ``L'' and ``H-SS'' prescriptions on the choice of the
IR cut-off parameters $\tilde{b}_{l}$ to the proton form factor entails
the modified HSP to be invalid. In view of these results, Li's analysis
of the proton form factor~\cite{Li93} seems to be seriously flawed.

A simple recipe to bypass the singular behavior of the integrand is
to ignore completely the evolution of the DA or to ``freeze'' it at any
(arbitrary) value larger than $\Lambda _{\text{QCD}}$. Hyer~\cite{Hye93}
suggested to take for the factorization scale
$\mu _{F}={\rm max}\left(1/b_{l}\right)$.
In this case, the $\tilde{\gamma}_{\text{max}}$
appears in (\ref{eq:kappa}) only if all three
$\tilde{b}_{l}$ tend to $1/\Lambda _{\text{QCD}}$ at once. But then at
least one of the $\exp[-s(\xi _{l},\tilde{b}_{l},Q)]$ drops to $0$
faster than any power of
$\ln\left(1/\tilde{b}_{l}\Lambda _{\text{QCD}}\right)$. Apparently,
Hyer's choice of the factorization scale avoids singularities of the
form (\ref{eq:sing}), but seems to us physically implausible.
Since he only presents numerical results for the proton form factor in
the time-like region, we cannot compare with his results directly.

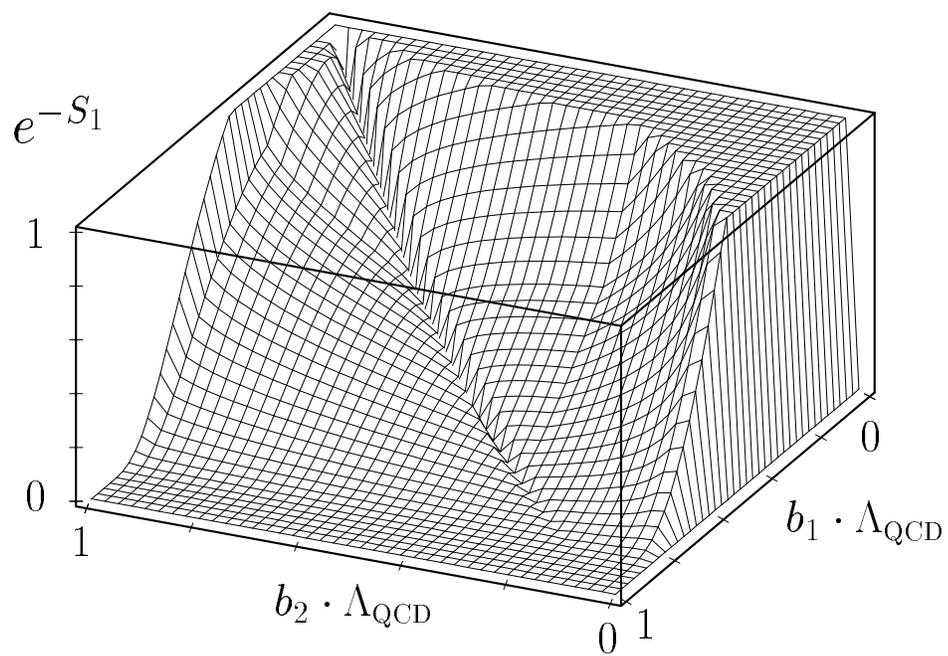
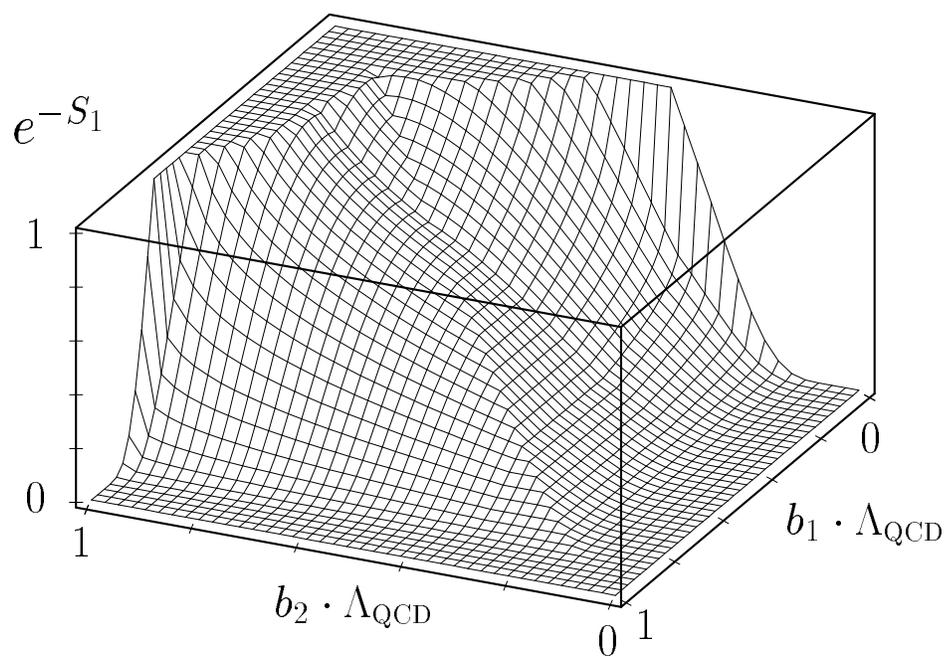
\begin{figure}
\caption{The Sudakov factor ${\rm e}^{-S_{1}}$ vs.
         $b_{1}\Lambda _{\text{QCD}}$ and $b_{2}\Lambda _{\text{QCD}}$
         evaluated at $Q=30\Lambda _{\text{QCD}}$, and
         $x_{1}=x_{1}^{\prime}=0.9$,
         $x_{2}=x_{3}=x_{2}^{\prime}=x_{3}^{\prime}=0.05$
         assuming a linear quark configuration
         ($\vec{b}_{1}$ and $\vec{b}_{2}$ are parallel to each other).
         The upper and lower figures correspond to the ``L''
         and ``MAX'' prescriptions, respectively.}
\label{fig:bcplots}
\end{figure}

Another option, and actually the one proposed in this work, is to use
a common IR cut-off not only for the evolution of the wave function
but also in the Sudakov exponent.
For a common cut-off $\tilde{b}$, the Sudakov factors always
cancel the $\alpha _{s}$-singularities; if, for a given $l$, we are in
the dangerous region,
$\xi _{l}<\sqrt{2}\Lambda _{\text{QCD}}/Q$,
$\tilde{b}\Lambda _{\text{QCD}} \to 1$,
at least one of the other two Sudakov functions lies in the region
$\xi _{l^{\prime}}>\sqrt{2}\Lambda _{\text{QCD}}/Q$,
$\tilde{b}\Lambda _{\text{QCD}} \to 1$ ($l^{\prime}\neq l$) and
therefore provides sufficient suppression, as outlined above.
In particular, we favor $\tilde{b}={\rm max}\{b_{l}\}$ as the optimum
choice (``MAX'' prescription), since it does not only lead to a regular
integral but also to a non-singular integrand.
The Sudakov factor ${\rm e}^{-S_{1}}$ subject to the ``L'' and ``MAX''
prescriptions is plotted for a specific quark configuration
in Fig.~\ref{fig:bcplots}. This figure makes it apparent that the
Sudakov factor in connection with the ``MAX'' prescription is
unencumbered by singularities in the dangerous soft regions.
As a consequence of the regularizing power of the ``MAX'' prescription,
the perturbative contribution to the proton form factor
(~\ref{eq:G_M(b)}) saturates in the sense that the results become
insensitive to the inclusion of the soft regions. A saturation as strong
as possible is a prerequisite for the self-consistency of the modified
HSP, as will be discussed in Sec.~V.

To demonstrate the amount of saturation, we calculate the proton form
factor through (\ref{eq:G_M(b)}), employing a cut-off procedure to the
$b_{l}$-integrations at a maximum value $b_{c}$.
In Fig.~\ref{fig:G_M(b_c)} the dependence of $G_{M}$ on $b_{c}$ for the
three choices, labeled: ``L'', ``H-SS'', and ``MAX'' is shown using,
for reasons of comparison with previous works, the COZ DA and ignoring
evolution.
[Evolution has been dispensed with to avoid the concomitant singularity
in $Q^{4}G_{M}$ as $b_{c}\Lambda _{\text{QCD}}\to 1$ when imposing the
``L'' and ``H-SS'' prescriptions.]
As one sees from the figure, the ``MAX'' prescription leads indeed to
saturation; the soft region
$b_{c}\Lambda _{\text{QCD}}>0.7$ does not contribute to the form factor
substantially. In fact, already $50\,\%$ of the result are obtained
from the regions with
$b_{c}\Lambda _{\text{QCD}}<0.48$. Note that $\alpha _{s}$ increases to
a value of $0.95$ at $b_{c}\Lambda _{\text{QCD}} \approx 0.48$.
This indicates that a sizeable fraction of the contributions to the
form factor is accumulated in the perturbative region.

Unfortunately, this saturation is achieved at the expense of
a rather strong damping of the perturbative contribution to the proton
form factor. Using the two other prescriptions (``L'' and ``H-SS'') and
ignoring evolution, we have found larger results for $G_{M}$, but
no indication for saturation: The additional contributions to the form
factor gained this way are accumulated exclusively in the soft
regions, i.e., for values of $b_{c}\Lambda _{\text{QCD}}$ near $1$.
These findings are in evident contradiction to Li's results
(figure 5 in~\cite{Li93}) for which an acceptable saturation has been
claimed.
On the other hand, we can qualitatively confirm the saturation behavior
of the proton form factor calculated by Hyer~\cite{Hye93} in the
time-like region.
Since we regard a saturation behavior as a stringent test for the
self-consistent applicability of pQCD, calculations which accumulate
large contributions from soft regions (large $b_{c}$) cannot be
considered as theoretically legitimate.

The r\^ole of the evolution effect subject to the ``MAX'' prescription
is also exhibited in Fig.~\ref{fig:G_M(b_c)}. It shows that the effect
of evolution is large, although finite, owing to the strong suppression
provided by the Sudakov factor. Note that according to our discussion in
Sec.~III, the factorization scale is $\mu _{F}=1/\tilde{b}$. The
significant feature of the evolution effect is that it tends to
neutralize the influence of the IR cut-off. Thus one obtains larger
values of the proton form factor at the expense of a slightly worse
saturation.

\section{NUMERICAL ANALYSIS}
\label{sec:numan}
In this section we give numerical results for the proton form factor.
In these calculations we throughout employ the ``MAX'' prescription
with evolution included, using $\Lambda _{\text{QCD}}=$180~MeV and
$\mu _{0}=1$~GeV.
Before proceeding with the presentation of our final results, let us
investigate the effect of including the intrinsic transverse momentum
in our calculations. The $k_{\perp}$-dependence of the proton wave
function effectively introduces a confinement scale in the formalism,
the importance of which may be appreciated by looking at
Fig.~\ref{fig:G_M(Q^2)}. This figure shows results, obtained for the
COZ DA without $k_{\perp}$-dependence and for two different values of
$\langle k^{2}_{\perp}\rangle ^{1/2}$.
To describe the intrinsic $k_{\perp}$-dependence, one can use
(\ref{eq:BLHM-Omega}) or, after transforming to the transverse
configuration space, (\ref{eq:FourierOmega}).
Notice that in Li's approach the Gaussian in (\ref{eq:FourierOmega})
has been replaced by unity. The oscillator parameter $a$ is determined
in such a way that either the normalization of the wave function
$P_{3q}$ is unity
(resulting into $\langle k^{2}_{\perp}\rangle ^{1/2}=271$~MeV for the
COZ DA),
or by inputing the value of the r.m.s. transverse momentum.
In the second case, we use a value of $600$~MeV (see the discussion
in Sec.~II), which implies $P_{3q}=0.042$. As can be seen from this
figure, the predictions for the form factor are quite different for
the three cases. The intrinsic $k_{\perp}$-dependence of the wave
function leads to further suppression of the perturbative contribution,
which becomes substantial if the r.m.s. transverse momentum is large.
On the other hand, this suppression is accompanied by an increasing
amount of saturation, since also the Gaussian (\ref{eq:FourierOmega})
suppresses predominantly contributions from the soft regions, viz., the
large $b$-regions. In contrast to the Sudakov factor, however, this
suppression is $Q$-independent. The interplay of the two effects,
Sudakov suppression and intrinsic transverse momentum,
leads to a different $Q$-behavior of the form factor depending on the
value of the r.m.s transverse momentum, as can be seen from
Fig.~\ref{fig:G_M(Q^2)}. The $Q$-dependence beyond $10$~GeV${}^{2}$
is rather weak, being approximately compatible with dimensional counting
(modulo logarithmic corrections). For very large values of $Q$ beyond
$1000$~GeV${}^{2}$ the three curves have approached each other within
$10\%$ accuracy. This happens when the Sudakov factor dominates the
Gaussian (\ref{eq:FourierOmega}) and selects those configurations with
small interquark separations.
In this region, which one may consider as the pure perturbative region,
the results for the form factor are independent of the confinement scale
introduced by the r.m.s. transverse momentum.

The penalty of the additional suppression of the perturbative
contribution caused by the Gaussian (\ref{eq:FourierOmega}) is
mitigated by the advantage that the perturbative contribution becomes
more self-consistent than by the Sudakov factor alone. This is indicated
in the enhanced amount of saturation with increasing r.m.s. transverse
momentum. Adapting the criterion of self-consistency, originally
suggested by Li and Sterman~\cite{LS92} for the pion case, namely that
$50\%$ of the results are accumulated at moderate values of the coupling
constant, say, $\alpha _{s}^{2}\leq 0.5$, we find self-consistency for
$Q^{2}\approx 7$~GeV${}^{2}$ (for the COZ DA).

Finally, in Fig.~\ref{fig:strip}, we demonstrate the effect of different
proton DAs on the form factor. To this end, we investigate a set of $45$
DAs~\cite{BS93,BS94}, which all respect the QCD sum-rule
constraints~\cite{COZ89a}. The results for the various
DAs---or more precisely
wave functions, since we include their intrinsic transverse momentum
dependence---obtained under the ``MAX'' prescription with evolution
included, form the shaded area shown in the figure. All wave functions
are normalized to unity and the corresponding r.m.s. transverse momenta
vary between $267$~MeV and $317$~MeV (see Table~1).
The theoretical form-factor predictions span a ``band'' congruent to the
``orbit'' of solutions found in~\cite{BS93,BS94}. The upper bound of the
``band'' corresponds to the DA COZ${}^{\text{up}}$, which
yields the maximum value of the form-factor
ratio $|G_{M}^{n}|/G_{M}^{p}=0.4881$ in the standard HSP.
The lower limit of the ``band'' is obtained using the DA ``low'' (sample
8 in~\cite{BS94}) with
$|G_{M}^{n}|/G_{M}^{p}=0.175$. Explicitly shown are
the results for the COZ DA~\cite{COZ89a}, its optimized version (with
respect to the sum-rule constraints) and the ``heterotic'' DA, recently
proposed by two of us in~\cite{SB93a}.
We note that the differences among these curves practically disappear
already at about $Q^{2}=80$~GeV${}^{2}$, despite the fact that these
amplitudes have distinct geometrical characteristics~\cite{Ste93a}.

Since the true valence Fock state probability is likely much smaller, or
invariably the r.m.s. transverse momentum larger than of order of
$300$~MeV, the ``band'' describes rather {\it maximal} expectations for
the (leading-order) perturbative contributions to the form factor; at
least for proton
wave functions of the type we utilize. Comparison with the experimental
data reveals that the theoretical predictions amount, at best, to
approximately $50\%$ of the measured values. This is the benchmark
against which we have to discern novelties and aberrations.
Closing this discussion we note that, since we are calculating only the
helicity-conserving part of the current matrix element
it is not obvious whether we
should compare the theoretical predictions with the data for the Sachs
form factor $G_{M}$ or the Dirac form factor $F_{1}$. Therefore we have
exhibited in Fig.~\ref{fig:strip} both sets of data~\cite{Roc82} for
comparison. Since the two sets of data differ by only $10\%$, our
conclusions concerning the smallness of the theoretical results remain
unaffected.

The various model wave functions led to self-consistency of the
perturbative contribution, i.e., $50\%$ of the results are accumulated
in regions where $\alpha _{s}^{2}\leq 0.5$, in the range of $Q^{2}$
between $6$ and $10$~GeV${}^{2}$.

\section{SUMMARY AND CONCLUSIONS}
\label{sec:con}
The objective of the present work has been to derive the proton magnetic
form factor within the modified version (Sec.~III) of the standard
Brodsky-Lepage HSP~\cite{LB80}, a scheme which takes into account
gluonic radiative corrections~\cite{CS81} in terms of transverse
separations. This is done by incorporating in the formalism the Sudakov
factor, calculated by Botts and Sterman~\cite{BS89}. There are already
some interesting applications of the modified
HSP~\cite{LS92,Li93,JK93,SS94,Hye93,GP94,CL93}.
The significant element of this type of analyses is that the
$\alpha _{s}$-singularities, arising from hard-gluon exchange and
evolution, can be cancelled without introducing free external
parameters. We emphasize that in contrast to pure phenomenological
recipies (e.g., the introduction of a gluon mass), the modified
HSP provides an explicit scheme how the IR protection of the ``bare''
$\alpha _{s}$ proceeds through gluonic radiation accumulated
in the Sudakov factor. Thus, in the modified HSP, one may conceive of
the (finite) IR-protected $\alpha _{s}$ as being the effective coupling.
By this procedure the potentially
dangerous soft regions of momenta are suppressed entailing also
a reduction of the perturbative contribution to the form factor.
While in the pion case~\cite{LS92}, it is fortunate that the
cancellation of the $\alpha _{s}$-singularities comes out naturally,
Li's approach to the proton form factor~\cite{Li93} leads to a lack of
complete cancellation of the $\alpha _{s}$-singularities (see Sec.~IV).
Without evolution of the proton wave function the emerging singularities
in (\ref{eq:G_M(b)}) are still integrable, but logarithmic corrections
due to evolution yield ultimately to uncompensated singularities.

On the grounds of our discussion, we are reasonably confident that Li's
treatment can be cured within the modified HSP. We suggest to use a
a common IR cut-off in the Sudakov exponents (\ref{eq:S}) and Sudakov
functions (\ref{eq:s}): viz., the maximum transverse separation.
This ``MAX'' prescription provides sufficient IR protection,
since even with evolution, the integrand in (\ref{eq:G_M(b)}) remains
finite. A significant feature of this treatment is that the proton form
factor saturates, i.e., it becomes insensitive to the contributions from
large transverse separations.
The other choices of the IR cut-off (``L'', ``H-SS''), we have
discussed, do not lead to saturation.

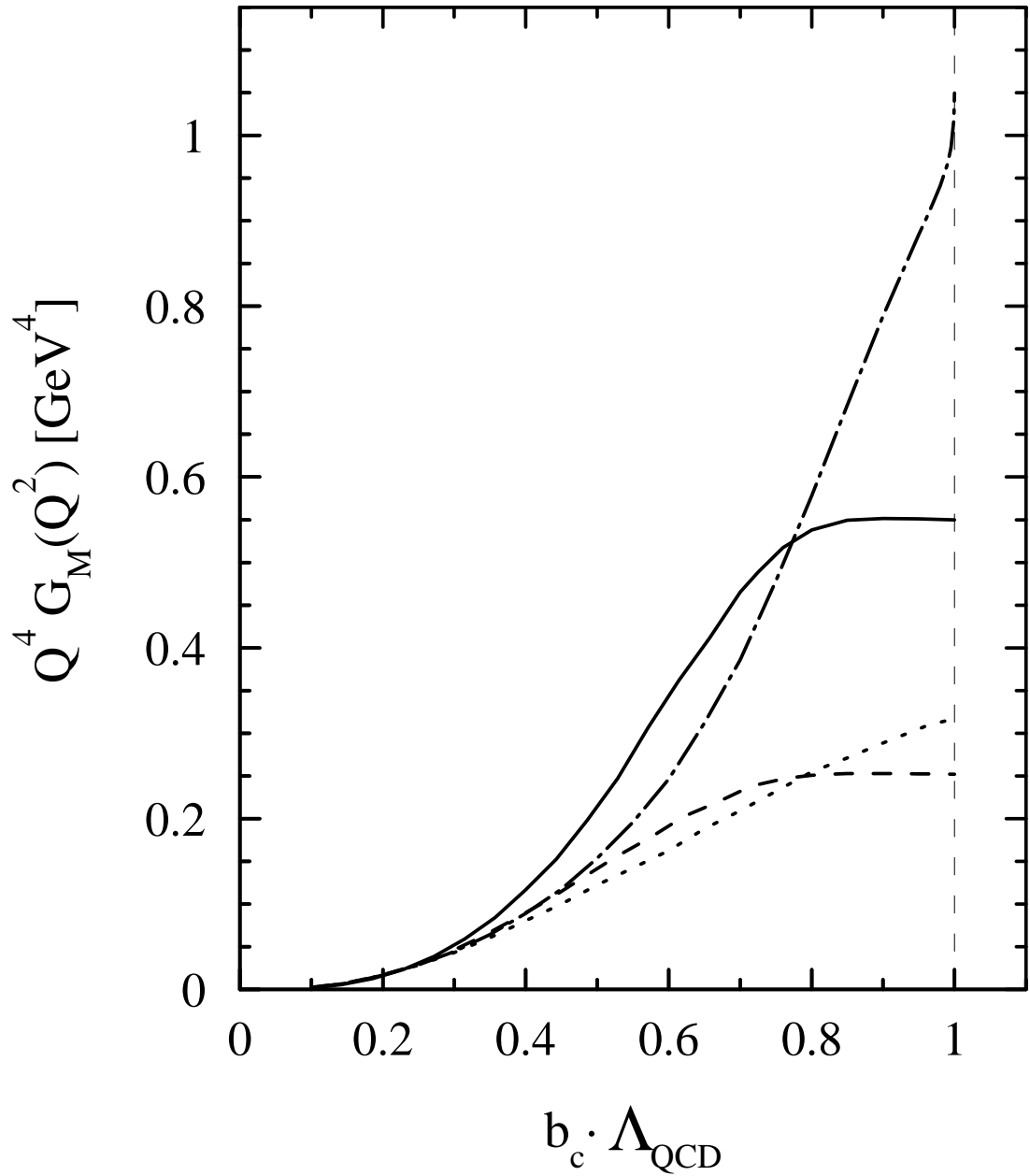
\begin{figure}
\caption{The proton magnetic form factor as a function of
         $b_{c}\Lambda _{\text{QCD}}$. The curves shown are
         obtained at $Q=30\Lambda _{\text{QCD}}$ for the COZ DA.
         The solid line corresponds to the ``MAX'' prescription
         including evolution.
         The dotted (dashed, dashed-dotted) line represents results
         using the ``H-SS'' (``MAX'', ``L'') prescription ignoring
         intrinsic $k_{\perp}$ and evolution.}
\label{fig:G_M(b_c)}
\end{figure}

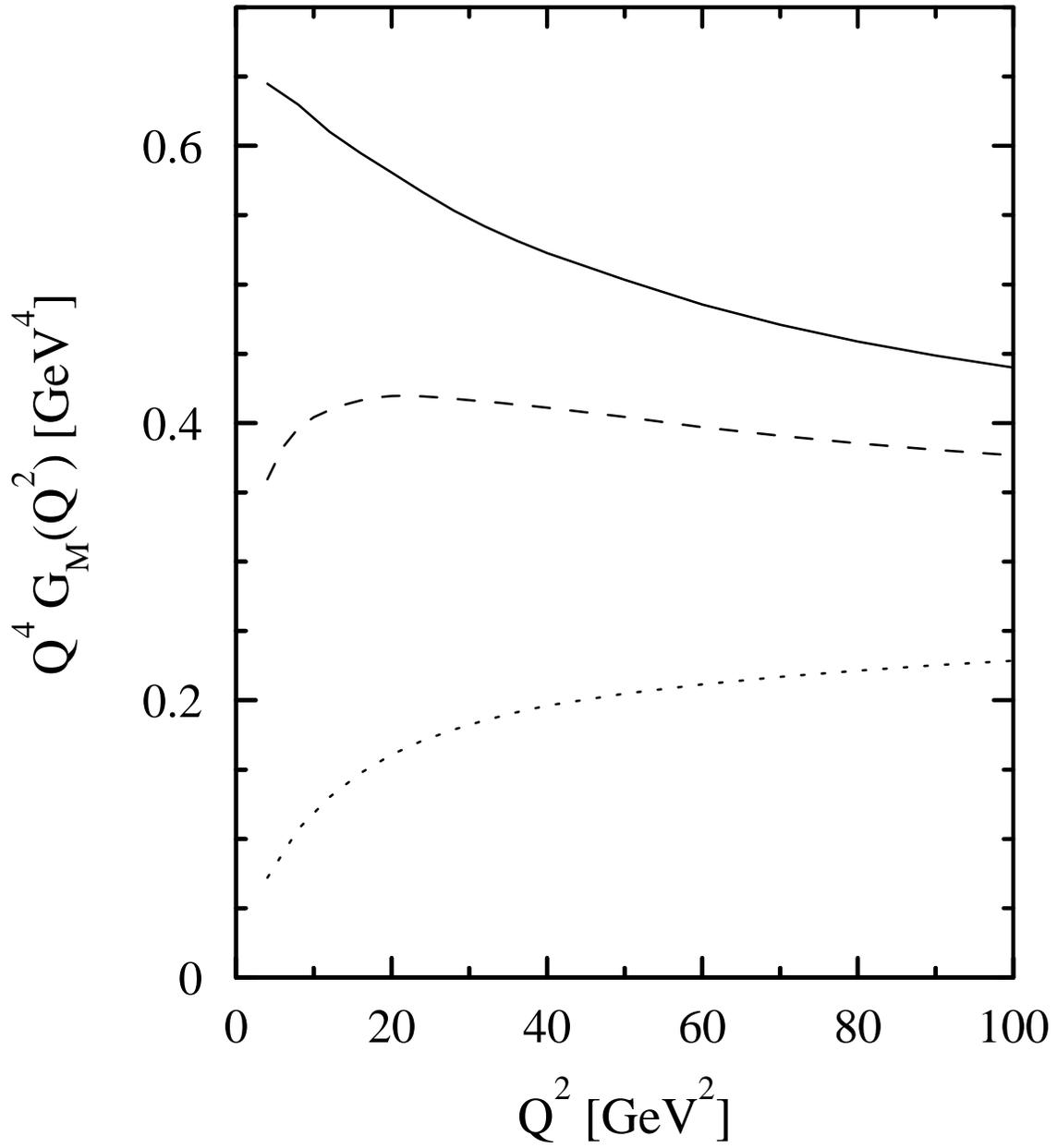
\begin{figure}
\caption{The influence of the intrinsic transverse momentum on the
         proton magnetic form factor. The curves shown are obtained for
         the COZ DA by imposing the ``MAX'' prescription including
         evolution. The solid line represents the results without
         $k_{\perp}$-dependence, whereas the dashed and dotted lines
         are obtained with $<k^{2}_{\perp}>^{1/2}=271$~MeV and
         $600$~MeV, respectively.}
\label{fig:G_M(Q^2)}
\end{figure}

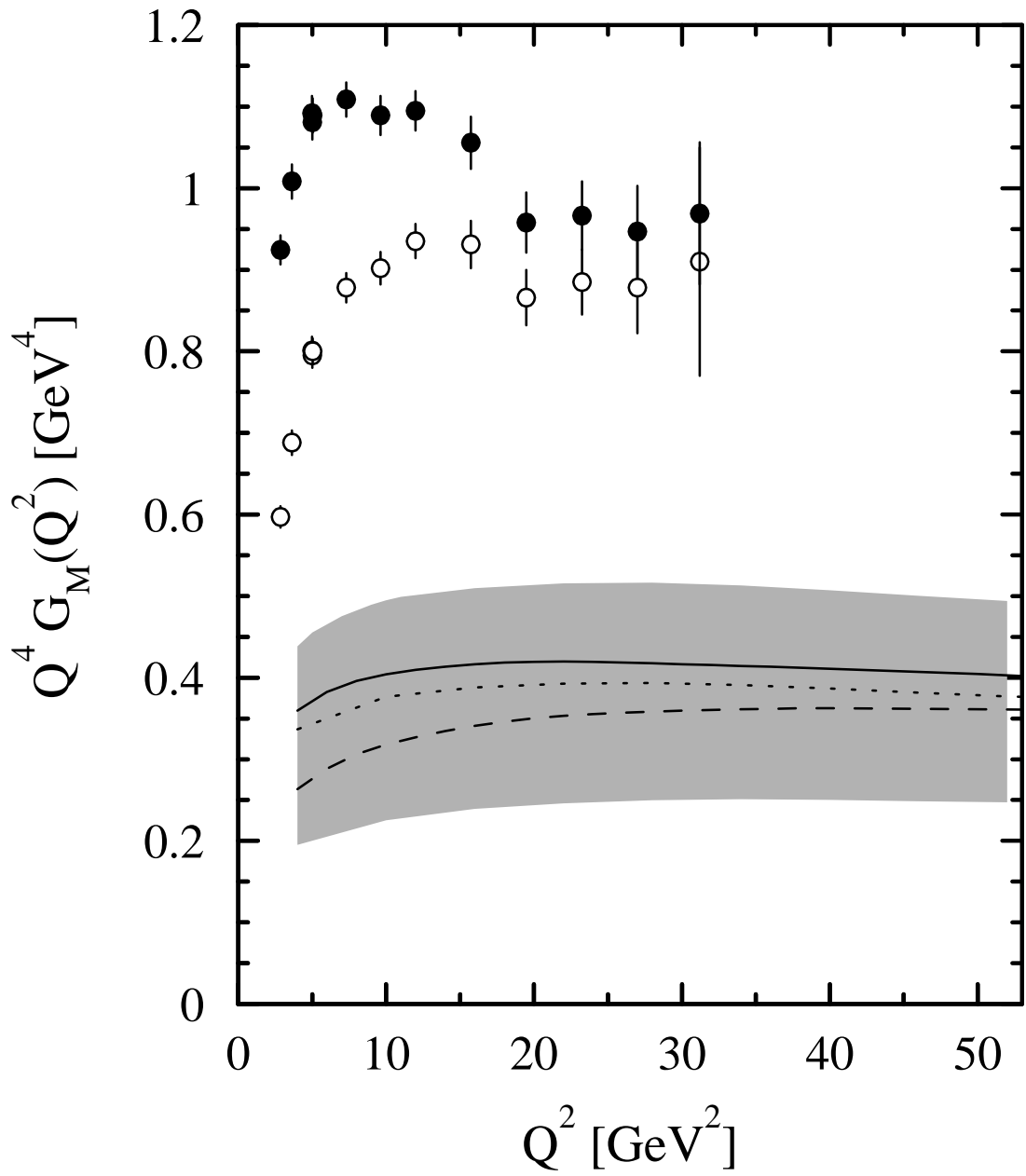
\begin{figure}
\caption[fig:strip]{The proton magnetic form factor vs. $Q^{2}$. Data
         are taken from \cite{Roc82}. The $G_{M}^{p}$ data are
         represented by black
         circles, whereas those for $F_{1}^{p}$ are indicated by open
         circles.
         The theoretical results are obtained using the ``MAX''
         prescription including evolution and normalizing the wave
         functions to unity. The shadowed strip indicates the range of
         predictions derived from the set of DAs determined in
         \cite{BS93,BS94} in the context of QCD sum rules (see text).
         The solid (dashed, dotted) line corresponds to the COZ
         (heterotic, optimized COZ) DA.}
\label{fig:strip}
\end{figure}

However, this reliable saturation and IR protection of the form factor
is achieved at the expense of a strong reduction of the perturbative
contribution to the form factor.
The damping of the proton form factor becomes even stronger if one takes
into account the intrinsic transverse momentum dependence of the proton
wave function (see Fig.~\ref{fig:G_M(b_c)} and Fig.~\ref{fig:G_M(Q^2)}).
This has been done by assuming a non-factorizing $x$ and
$k_{\perp}$-dependence of the wave function of the
Brodsky-Lepage-Huang-Mackenzie~\cite{BHLM83} type and fixing the value
of $\langle k^{2}_{\perp}\rangle ^{1/2}$ either via the valence quark
probability $P_{3q}$ or by inputing the value $600$~MeV by
hand~\cite{SS94}.
A remarkable finding is that the form factor calculated within the
modified HSP, appropriately extended to include the intrinsic transverse
momentum of the proton wave function, shows only a mild dependence on
the particular model DA.

The perturbative contribution to the form factor becomes self-consistent
in all cases for momentum transfers larger than $6$ to $10$~GeV${}^{2}$.
The actual value of the onset of self-consistency depends on the
particular wave function and the r.m.s. transverse momentum chosen.
Self-consistency is defined such that $50\%$ of the result are
accumulated in regions where $\alpha _{s}^{2}$ is smaller than $0.5$
(Sec.~V).

Comparing our theoretical results with the data, it turns out that
they fall short by at least $50\%$. This is true not only for the COZ DA,
(which we have exemplarily used to facilitate comparison with previous
works) but actually for the whole spectrum of amplitudes determined
in \cite{BS93,BS94} and found to comply with the COZ sum-rule
requirements. Depending on the
actual value of the r.m.s. transverse momentum, the reduction of the
perturbative contribution may be even stronger than $50\%$.

The fact that in all considered cases the self-consistently calculated
perturbative contribution to the proton form factor fails to reproduce
the existing data, is perhaps a signal that soft contributions (higher
twists) not accounted for so far by the modified HSP should be included.
Such contributions comprise, e.g., improved and/or more complicated wave
functions, orbital angular momentum, higher Fock components, quark-quark
correlations (diquarks), radiative corrections to the quark and gluon
condensates, quark masses, etc. Also remainders of genuine soft
contributions, like
vector-meson-dominance terms or the overlap of the soft parts of the
wave functions (Feynman contributions), may still be large at accessible
momentum transfers. The rather large value of the Pauli form factor
$F_{2}$ around $10$~GeV${}^{2}$, as found experimentally~\cite{Bos92},
indicates that sizeable higher-twist contributions still exist in
that region of momentum transfer~\cite{Kro93}. One may suspect
similar or even larger higher-twist contributions to the helicity
non-flip current matrix element controlling $F_{1}$ and $G_{M}$.
Large (perturbative) higher-order corrections to the hard-scattering
amplitude cannot be excluded as well, since their size has not yet been
estimated. In analogy to the Drell-Yan process, these corrections may be
condensed in a K-factor multiplying the leading-order perturbative
result. However, with our
choice of the renormalization scale, the K-factor is expected to be
close to unity. At least for the case of the pion form factor,
calculations of the K-factor to one-loop order exist~\cite{Fie81,DR81},
which indicate that choosing the renormalization scale analogously to
ours, the value of the K-factor is indeed close to unity.

In conclusion we note that it was not our primary aim to use
the modified HSP to obtain best agreement with the data, although from
our point of view this scheme represents a decisive step towards a
deeper understanding of the electromagnetic form factors.
In the present work the focus has been placed on theoretical problems,
overlooked previously.

\acknowledgements
We would like to thank James Botts for useful discussions and Stan
Brodsky, Thomas Hyer and Hsiang-nan Li for private communications.

\newpage 


\newpage 

\begin{table}

\caption[table]{Expansion coefficients for selected DAs, taken from
         \cite{COZ89a} and \cite{BS93,BS94}. Our notation is adopted
         from \cite{Ste89}. The $\{\tilde{\gamma}_{n}\}$ are related to
         the anomalous dimensions of trilinear quark operators. The
         associated r.m.s transverse momentum and oscillator
         parameter for each model wave function, normalized via
         $P_{3q}=1$, are shown.}

\vspace{0.5cm}
\begin{tabular}{lrrrrrr}
 $\;n$ & $\tilde \gamma_n$
   & $B_n$(COZ$^{\mbox{\scriptsize up}}$) & $B_n$(COZ)
   & $B_n$(COZ$^{\mbox{\scriptsize opt}}$)
   & $B_n$(het) & $B_n$(low) $\;$ \cr
 \tableline
 $\;1$ & 20/9  &  3.2185  &  3.6750  &  3.5268 &  3.4437
       &  4.1547 $\;$\cr
 $\;2$ & 24/9  &  1.4562  &  1.4840  &  1.4000 &  1.5710
       &  1.4000 $\;$\cr
 $\;3$ & 32/9  &  2.8300  &  2.8980  &  2.8736 &  4.5937
       &  3.3756 $\;$\cr
 $\;4$ & 40/9  & -17.3400 & -6.6150  & -4.5227 & 29.3125
       & 26.1305 $\;$\cr
 $\;5$ & 42/9  &  0.4700  &  1.0260  &  0.8002 & -0.1250
       & -0.5855 $\;$\cr
 \tableline \tableline
 \multicolumn{2}{l}{$\langle k_{\perp}^2\rangle ^{1/2}$
      [MeV]} & 271 & 271 & 267 & 317 & 299 $\;$\cr
 \multicolumn{2}{l}{$a$ [GeV$^{-1}$]} & 0.9893 & 0.9939 & 1.0069 &
      0.8537 & 0.9217 $\;$\cr
\end{tabular}
\end{table}

\newpage  

\end{document}